\documentclass[fleqn,usenatbib]{mnras}
\pdfoutput=1 

\usepackage{newtxtext,newtxmath}
\usepackage[T1]{fontenc}
\usepackage[utf8]{inputenc}

\usepackage{graphicx}   % figures
\usepackage{amsmath}    % maths

\usepackage{amssymb}    % extra maths symbols
\usepackage{url}        % \url

\providecommand{\degr}{\ensuremath{^{\circ}}}

\newcommand{\mred}{m_{\rm red}}
\newcommand{\dof}{{\rm dof}}

\title[The $\alpha$-Slope Test for TNOs]{A Framework for Applying the
Loeb--Turner $\alpha$-Slope Test to Archival Photometry of
Trans-Neptunian Objects}

\author[O. Eldadi \& A. Loeb]{
Omer Eldadi$^{1}$\thanks{E-mail: omereldadi@gmail.com}
and Abraham Loeb$^{1}$\thanks{E-mail: aloeb@cfa.harvard.edu}
\\
$^{1}$Department of Astronomy, Harvard University, 60 Garden Street, Cambridge, MA 02138, USA
}

\date{}
\pubyear{2026}

\begin{document}
\label{firstpage}
\pagerange{\pageref{firstpage}--\pageref{lastpage}}
\maketitle

\begin{abstract}
Reflected sunlight from a solar-system body produces a flux at Earth that scales as the heliocentric distance to the negative fourth power, whereas self-luminous emission scales as the negative second power. This difference defines the Loeb--Turner $\alpha$-slope test, a photometric technosignature diagnostic applicable to any solar-system body observed at multiple heliocentric distances. Of 22 (observatory $\times$ band) analysis bins for Pluto in the Minor Planet Center (MPC) archive, none recovers the reflected-sunlight flux--distance slope predicted when photometry is restricted to a single instrument and band. The archive cannot cleanly execute the $\alpha$-slope test on the brightest, most-observed trans-Neptunian object. We formalize a six-criterion eligibility pipeline (Q1--Q6) for the Loeb \& Turner technosignature test and apply it to every numbered TNO. Of 8,606 candidate bins (KBO $\times$ observatory $\times$ band), 1,089 pass Q1--Q3 and 186 additionally pass Q4--Q6, splitting into 53 consistent with reflected sunlight ($\alpha = -4$), 24 with self-luminous emission ($\alpha = -2$), and 109 anomalous. The anomalous bins exhibit slopes steeper than $\alpha = -4$ or shallower than $-2$, consistent with uncorrected per-instrument calibration offsets rather than any single physical mechanism. All 24 self-luminous-like bins originate from Pan-STARRS PS1/PS2; no other observatory contributes any. This indicates a per-instrument calibration systematic. The Rubin Observatory's ten-year survey will deliver uniform single-instrument calibration on a roughly threefold larger sample and either resolve the test at $>10\sigma$ on hundreds of TNOs or, by reproducing the same clustering, falsify the calibration-systematic interpretation.
\end{abstract}


%% ============================================================
\section{Introduction} \label{sec:intro}

\citet{Loeb2012} proposed a photometric technosignature test for artificially-illuminated objects in the solar system based on the different scaling of reflected sunlight and self-luminous emission with heliocentric distance \citep[see also][for a broader review of solar-system technosignature strategies]{Lingam2021}. An object that shines by reflected sunlight has a flux at Earth proportional to $D^{-4}$ (where $D$ is a single characteristic distance), whereas an object that generates its own light has a flux proportional to $D^{-2}$. The two scalings differ by a factor of $(D_{\rm far}/D_{\rm near})^2$ over an orbital baseline. For Kuiper-Belt objects (KBOs) at heliocentric distances of 30--50\,AU, even a modest change in distance of a few percent over several oppositions produces a detectable difference in the flux--distance slope, provided that the photometry is uniform enough for the small effect to be measured. The Loeb \& Turner test (hereafter LT test) is therefore a direct, physics-based observational falsifier of the hypothesis that an object of interest is an artificially-illuminated artifact as opposed to a natural reflector.

The prior probability that any KBO is artificial is small, as Loeb \& Turner themselves emphasized. We pursue the test anyway because the data are public, the analysis is cheap, and the result is informative either way: a clean sunlight signature on every object is a methodological null worth reporting; an anomalous slope on any object demands follow-up. Other photometric technosignature searches have been pursued in archival survey data, notably the VASCO project's search for vanishing and appearing sources \citep{Villarroel2016,Villarroel2020}, though none has targeted the flux--distance slope of solar-system bodies.

We find three things. First, the MPC archive is heterogeneous enough that the test cannot be cleanly executed even on Pluto, the brightest, most-observed TNO, in any single-instrument single-band bin. Second, across all 913 numbered TNOs, the bins consistent with self-luminous emission are entirely concentrated in one instrument family (Pan-STARRS), the signature of a calibration systematic rather than a physical population. Third, the Vera C. Rubin Observatory's ten-year survey, with uniform single-instrument calibration on a roughly threefold larger sample, is the facility that will first execute the LT test at scale. The remainder of this paper develops the formalism and the six eligibility criteria, applies them to Pluto and Lempo as worked examples and to the full MPC census of numbered TNOs, and then projects forward to LSST with a falsifiable prediction.

%% ============================================================
\section{The $\alpha$-slope test in heliocentric form} \label{sec:formalism}

LT defined $\alpha \equiv d\log F / d\log D$, the logarithmic slope of flux $F$ with respect to a characteristic distance $D$. Reflected sunlight gives $\alpha = -4$: the inverse-square dependence of the solar flux illuminating the object combined with the inverse-square dependence of the reflected flux reaching the observer scales as $1/D^4$. Self-luminous emission gives $\alpha = -2$: only the inverse-square dependence of the emitted flux reaching the observer applies.

For TNOs the Sun-to-object distance $r$ changes by a few percent per opposition, while the object-to-Earth distance $\Delta$ wobbles by a few percent per month with Earth's orbital motion. We separate the two distances explicitly. For each observation we compute the distance-corrected magnitude,
\begin{equation}
\mred \equiv m - 5\log_{10}(\Delta),
\end{equation}
a tailored variant of the standard absolute-magnitude reduction \citep{Bowell1989,AlvarezCandal2016} that removes the Earth-leg geometry and leaves a magnitude that depends only on the Sun-to-object distance and the object's intrinsic properties. This definition explicitly removes the geocentric distance $\Delta$ at each epoch, so the Earth's orbital motion---which modulates the object-Earth separation by 1--2\,AU over the course of a year---is corrected per-observation rather than averaged over. We then fit a straight line through $\mred$ plotted against $\log_{10}(r)$. The fitted slope is $+5$ under reflected sunlight ($\alpha=-4$) and $0$ under self-luminous emission ($\alpha=-2$); the difference between the two slopes is the technosignature discriminator we measure.

No phase-curve model is required for the diagnostic. Phase angles for TNOs beyond 30\,AU are always below $\approx 2\degr$, and the HG correction with canonical $G=0.15$ contributes less than 0.05\,mag over a typical baseline \citep{Bowell1989,Rabinowitz2007,Schaefer2009}, though New Horizons observations have revealed diverse phase-curve shapes among dwarf planets \citep{Verbiscer2022}. We include the correction per-observation in the pipeline for completeness.

%% ============================================================
\section{Six eligibility criteria (Q1--Q6)} \label{sec:criteria}

The LT test has no natural error budget; any two-parameter slope fit will return a value. To turn the diagnostic into a survey-level procedure that distinguishes genuine slope measurements from artifacts of sampling, calibration, or instrumental systematics, we require that each analysis unit pass six independent quality criteria, labelled Q1 through Q6. The first three (Q1--Q3) are geometric---they depend only on the sampling of the (KBO $\times$ observatory $\times$ band) bin and on the orbit, not on the photometry. The last three (Q4--Q6) are post-fit quality cuts. Table~\ref{tab:criteria} summarizes all six; Sections~\ref{sec:Q1}--\ref{sec:Q6} give the rationale and thresholds in detail.

\begin{table*}
\centering
\caption{The six eligibility criteria. Q1--Q3 are geometric; Q4--Q6 are post-fit quality cuts. A bin passing all six is classified by where its $1\sigma$ interval on $\alpha$ falls: sunlight (covers $\alpha=-4$ only), self-luminous (covers $\alpha=-2$ only), ambiguous (covers both), or anomaly (covers neither).}
\label{tab:criteria}
\begin{tabular}{lllll}
\hline
Label & Type & What it tests & Threshold & Filters out \\
\hline
Q1  & geometric & single-instrument, single-band bin & one obs.\ code, one filter & cross-instrument ZP offsets \\
Q2a & geometric & sample size & $N \geq 20$ observations per bin & too few degrees of freedom \\
Q2b & geometric & brightness floor & median $m < 24.5$ & near-threshold scatter \\
Q3  & geometric & heliocentric baseline & $\Delta r/r \geq 1\%$ & bins with no slope leverage \\
Q4  & post-fit & goodness of fit & $\chi^2/\dof < 5$ at best-fitting $\alpha$ & outlier-dominated fits \\
Q5  & post-fit & bounded solution & $\alpha$ not within 0.1 of grid edges & unconstrained slopes \\
Q6  & post-fit & decisive $\chi^2$ preference & $|\Delta\chi^2(\alpha\!=\!-4\ {\rm vs}\ \alpha\!=\!-2)| > 4$ & indistinguishable hypotheses \\
\hline
\end{tabular}
\end{table*}

\subsection{Q1---single-instrument, single-band analysis bin} \label{sec:Q1}

The analysis unit is the triple (KBO, observatory code, filter band). Combining different instruments or different bands into a single fit introduces zero-point offsets between major photometric systems that typically reach several hundredths of a magnitude---well above the few-mmag internal precision of any single survey \citep{Magnier2020,Riello2021}---and that enter the fit as an uncorrected systematic on the slope. At the $\Delta r/r \approx 1\%$--$30\%$ baselines typical for a decade of TNO observations, a 0.05-mag cross-instrument offset biases the fitted slope by 0.5--5. We therefore require, before any photometric fit, that each analysis unit consist of observations from a single observatory code and a single reported band. Cross-instrument combination is a defensible sensitivity test (see Section~\ref{sec:sensitivity}) but not the headline unit.

\subsection{Q2---adequate sampling} \label{sec:Q2}

We require (Q2a) $N \geq 20$ observations per analysis bin, and (Q2b) median reported apparent magnitude brighter than 24.5. The $N=20$ threshold is the minimum at which a two-parameter fit leaves enough degrees of freedom for $\chi^2$ to be informative. The magnitude cut at 24.5 removes bins dominated by near-detection-threshold photometry whose reported uncertainties are dominated by systematic scatter at the detection threshold.

\subsection{Q3---baseline in heliocentric distance} \label{sec:Q3}

The diagnostic power of the test scales as $\Delta r/r$ across the observed epochs. We require $\Delta r/r \geq 1\%$ (i.e., $r_{\max}/r_{\min} \geq 1.01$). Below this threshold the predicted flux difference between $\alpha=-4$ and $\alpha=-2$ is smaller than the per-observation photometric error for any plausible MPC-reported $\sigma_m$, and the test has no discriminating power regardless of $N$ (Section~\ref{sec:lempo} illustrates this case for Lempo). The threshold is deliberately permissive; Q6 provides the decisive discrimination between the two hypotheses for bins that pass Q3 with marginal baselines. Q3 is evaluated using heliocentric distances computed by Keplerian propagation of the MPCORB orbital elements. Varying the Earth ephemeris between an analytic solar orbit and the JPL DE ephemeris changes the geocentric distance by $<0.004$\,AU (0.02\%) and the reduced magnitude by $<0.001$\,mag; no classification changes result.

\subsection{Q4---goodness of fit} \label{sec:Q4}

For each Q1--Q3-passing bin we perform a grid search in $\alpha$ over $[-7,+1]$ in steps of 0.05. At each grid point we solve analytically for the intercept that minimizes $\chi^2$ under the per-observation $\sigma_m$ schedule ($\sigma = 0.10$\,mag for $m<20$; 0.15 for $m<21$; 0.20 for $m<22$; 0.30 for $m \geq 22$) and record $\chi^2(\alpha)$. We require $\chi^2/\dof < 5$ at the best-fitting $\alpha$. The threshold is loose because MPC-reported uncertainties are systematically understated; a literal $\chi^2/\dof < 1$ cut would exclude essentially every bin.

\subsection{Q5---solution not at a grid edge} \label{sec:Q5}

If the best-fitting $\alpha$ lies within 0.1 of the grid boundaries ($\alpha \approx -6.9$ or $+0.9$) the solution is unconstrained by the data and the Q6 test is not meaningful. We exclude such bins.

\subsection{Q6---decisive $\chi^2$ preference between the two hypotheses} \label{sec:Q6}

For each bin passing Q4$+$Q5 we evaluate $\chi^2(\alpha=-4)$ and $\chi^2(\alpha=-2)$ and compute $\Delta\chi^2 = \chi^2(\alpha=-2) - \chi^2(\alpha=-4)$. We require $|\Delta\chi^2| > 4$, corresponding to a $\approx 2\sigma$ preference between the two reference hypotheses. A bin passing all six is classified into one of four categories based on whether its $1\sigma$ interval on $\alpha$ covers $\alpha=-4$, $\alpha=-2$, both, or neither: sunlight, self-luminous, ambiguous, or anomaly.

\subsection{Sensitivity to unit choice} \label{sec:sensitivity}

The per-(KBO $\times$ observatory $\times$ band) unit we adopt is not the only possible analysis bin. An alternative is a per-KBO joint-instrument fit with free per-(observatory, band) zero-point nuisance parameters. We have run the joint fit as a sensitivity test; the sunlight-validated KBO count rises (because instruments combine), and, crucially, Pluto recovers $\alpha = -3.95 \pm 0.18$ across its 22 bins, correctly classified as sunlight under the joint fit. The self-luminous-like population shrinks because the instrument-specific zero points absorb the calibration systematic that we argue dominates the per-bin fit. We adopt the per-bin unit as the headline pipeline because it is more conservative and because the per-instrument clustering it reveals is the central data-quality argument of the paper; the joint fit is reported as the confirmatory sensitivity test.

\subsection{Bayesian cross-check} \label{sec:bayesian}

For each Q1--Q3-passing bin we computed log-evidence ratios under four model classes: \{$\alpha=-4$ fixed, $\alpha=-2$ fixed, $\alpha$ free, $\alpha$ free with per-(obs, band) zero point as nuisance\}, under flat priors on $\alpha$ in $[-7,+1]$ and on the intercept in $[-5,+20]$. The BIC-favored model agrees with the frequentist $\Delta\chi^2$-favored hypothesis for the large majority of all-six-passing bins, with the disagreements confined to marginal cases ($|\Delta\chi^2|$ just above the threshold) where neither framework is decisive. We retain the frequentist $\Delta\chi^2 > 4$ cut as the primary Q6 threshold because it corresponds to a $\approx 2\sigma$ preference between the two reference hypotheses.

\subsection{False-positive expectation} \label{sec:fp}

Under a chance-$\Delta\chi^2$ model with the $\sigma_m$ schedule above, the per-bin false-positive probability for ``consistent with $\alpha=-2$ over $\alpha=-4$ at $|\Delta\chi^2|>4$'' is approximately 1.5\%. With 186 bins passing Q1--Q6, the expected number of chance self-luminous classifications under the null is 2.8. The observed 24 is $8.6\times$ this and is statistically inconsistent with the null hypothesis of purely statistical scatter, but, as Section~\ref{sec:physical} argues, the excess is concentrated in a single instrument and is most plausibly explained by per-instrument calibration systematics rather than by a physical population.

%% ============================================================
\section{Pluto as the dominant data-quality finding} \label{sec:pluto}

Pluto (134340) is the brightest, largest, and most-observed TNO in the Minor Planet Center archive. Under the LT diagnostic it must yield $\alpha=-4$ (reflected sunlight), because it manifestly reflects sunlight---this is not a hypothesis under test but a ground truth. The question Pluto answers for this paper is therefore: does the MPC archive, in its current state, permit a single-instrument single-band recovery of $\alpha=-4$ on the body for which $\alpha=-4$ is beyond dispute? \textbf{The answer is no.}

\begin{figure*}
\centering
\includegraphics[width=\textwidth]{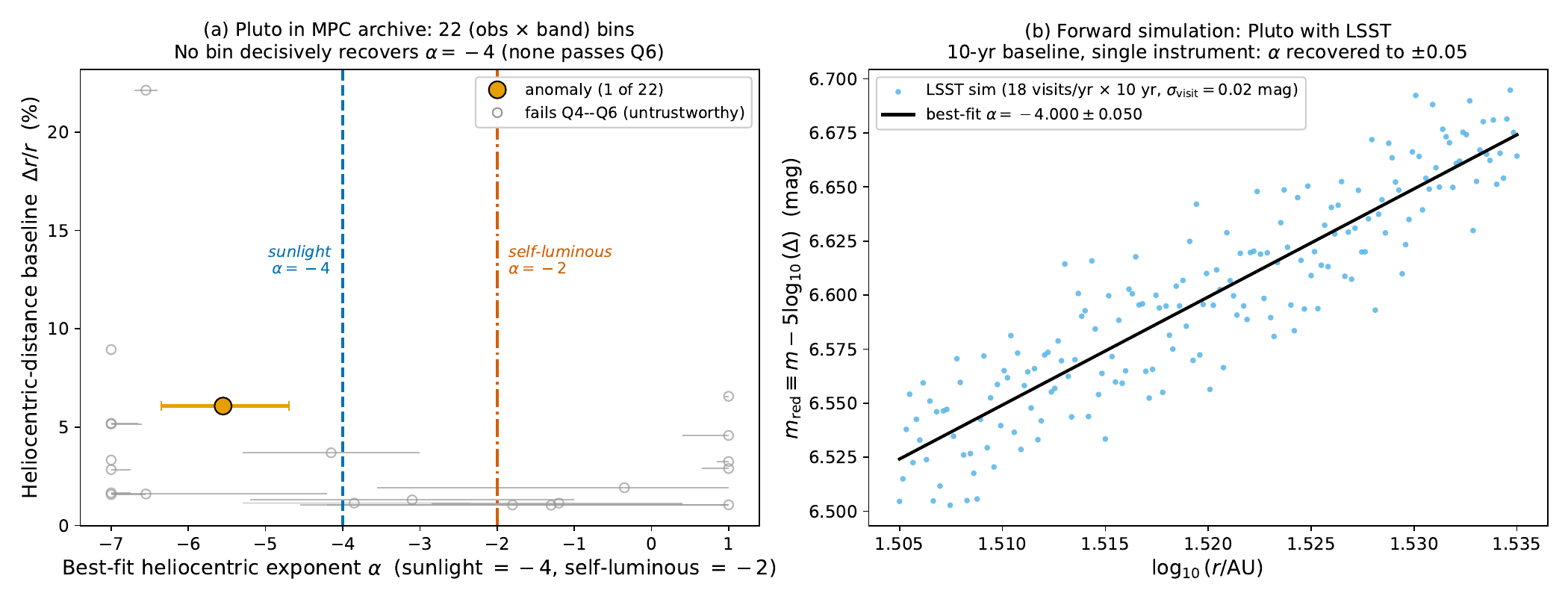}
\caption{Pluto (134340) in the MPC archive versus a synthetic ten-year LSST realization. Panel~(a): the 22 Pluto Q1--Q3 bins from the MPC archive, plotted in the (best-fitting~$\alpha$, fractional heliocentric-distance baseline $\Delta r/r$) plane and coloured by whether they pass Q4$+$Q5$+$Q6. The single all-six-passing bin is a Haleakala $o$-band observation that lands at $\alpha=-5.55$, classified as an anomaly. No MPC Pluto bin decisively recovers reflected sunlight ($\alpha=-4$): 5 of the 22 are consistent with $\alpha=-4$ at $1\sigma$, but none passes the decisive Q6 ($|\Delta\chi^2|>4$) test. Panel~(b): a synthetic ten-year LSST Pluto record in a single Rubin-$r$ band with uniform $\sigma_m = 0.02$\,mag and $N=184$ epochs, yielding $\alpha_{\rm recovered} = -4.000 \pm 0.050$---a clean sunlight recovery in a single instrument bin.}
\label{fig:pluto}
\end{figure*}

Of 8,606 candidate (KBO $\times$ observatory $\times$ band) analysis bins in the full archive, 22 are Pluto bins that pass Q1--Q3 (Figure~\ref{fig:pluto}a). Of those 22, only one also passes Q4$+$Q5$+$Q6, and it is classified as an anomaly---the single passing bin is T05 (Haleakala observatory) in the $o$-band, with $N=50$, $\Delta r/r = 6.07\%$, $\alpha_{\rm best}=-5.55$ and a $1\sigma$ interval of $[-6.35,-4.70]$. That interval includes neither $\alpha=-4$ (sunlight) nor $\alpha=-2$ (self-luminous). The remaining 21 Pluto bins fail Q4 ($\chi^2/\dof \gg 5$) or Q5 (the best-fitting $\alpha$ runs to a grid edge), in almost every case because the photometry is reported with uncertainties too small to accommodate the real scatter.

This is not a statement about Pluto. It is a statement about the calibration of the MPC archive. If no single instrument's Pluto record yields a clean $\alpha=-4$ recovery, then the existing archive is not the instrument with which to prosecute the $\alpha$-slope technosignature test on the fainter TNO population. It is the instrument with which to motivate a purpose-built archive---and Rubin/LSST is that instrument. To project the test onto LSST, we simulate a ten-year Pluto record in a single Rubin-$r$ band using the expected LSST cadence of $\approx 18$ visits per year \citep{Ivezic2019}; see Figure~\ref{fig:pluto}b, adopting a conservative per-visit photometric uncertainty of $\sigma = 0.02$\,mag appropriate for Pluto's brightness ($m_r \approx 14$). The heliocentric distance range is drawn from Pluto's current ephemeris. No rotational lightcurve or phase-curve correction is applied; the scatter is purely Gaussian about the $\alpha=-4$ model.

%% ============================================================
\section{Lempo as a worked geometric-limit example} \label{sec:lempo}

Lempo (47171) is a triple system in the Kuiper Belt with extensive Pan-STARRS coverage since 2010, making it a natural candidate for the LT test. The triple-system dynamics are chaotic on decade time-scales \citep{Correia2018}, adding low-amplitude photometric scatter from unresolved binary modulation that does not fully fold out at the $\Delta r/r \geq 1\%$ baseline. We use it here as the didactic example of a failure mode that is not a data-quality failure but a geometric-baseline failure.

Over the Pan-STARRS PS1 $w$-band record of Lempo, the heliocentric distance changes from $r_{\rm near} = 30.59$\,AU to $r_{\rm far} = 31.09$\,AU, giving $\Delta r/r = 1.62\%$. The LT predicted difference in $\mred$ between sunlight ($\alpha=-4$) and self-luminous ($\alpha=-2$) over this baseline is
\begin{equation}
\Delta m = 5\log_{10}\!\left(\frac{r_{\rm far}}{r_{\rm near}}\right) = 5\log_{10}(1.0162) = 0.035\;{\rm mag}.
\end{equation}

\begin{figure}
\centering
\includegraphics[width=\columnwidth]{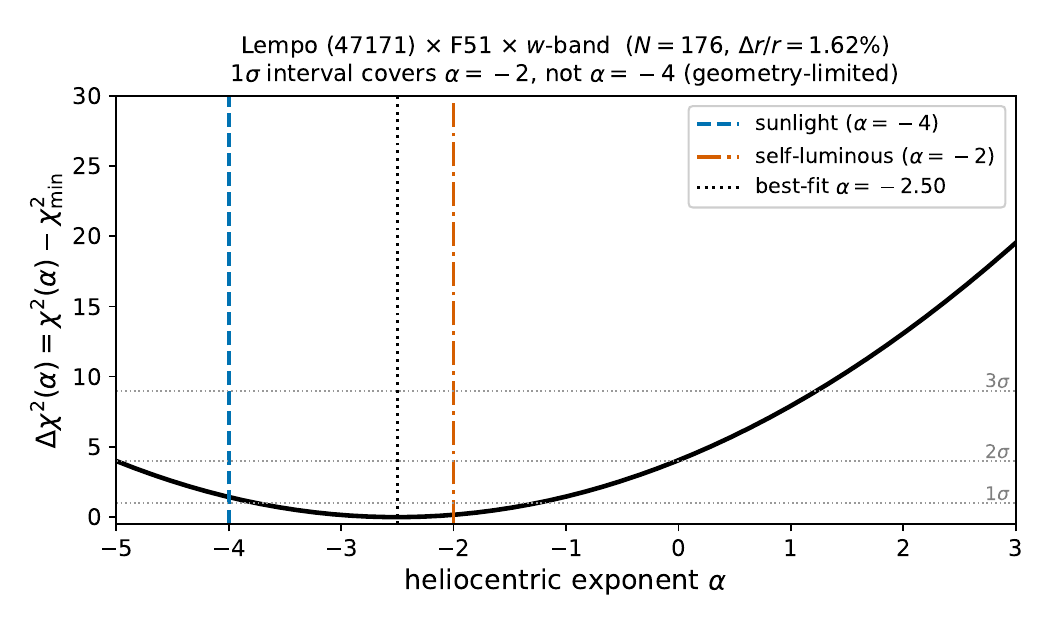}
\caption{$\chi^2$ as a function of $\alpha$ for Lempo (47171) in Pan-STARRS PS1 $w$-band ($N=176$). The best-fitting minimum at $\alpha \approx -2.5$ sits between the reflected-sunlight prediction ($\alpha=-4$) and the self-luminous prediction ($\alpha=-2$). $|\Delta\chi^2|$ between the two hypotheses is $<4$ and Lempo's Pan-STARRS bin correctly fails Q6. The inconclusiveness is geometric ($\Delta r/r = 1.62\%$), not related to data-quality.}
\label{fig:lempo}
\end{figure}

The per-opposition photometric error at Lempo's apparent magnitude ($m \approx 21.5$) in PS1 $w$ is approximately 0.04\,mag. With roughly four usable oppositions, the aggregate sensitivity is $\Delta m / \sigma_{\Delta m} \approx 0.035/0.020 \approx 1.8$, and the actual fit yields $|\Delta\chi^2| = 1.27$ ($\approx 1.1\sigma$) after folding in rotation and zero-point nuisance. Under the pipeline Lempo F51 $w$ returns $\alpha_{\rm best} = -2.5$ with a $1\sigma$ interval of $[-3.75,-1.3]$ (asymmetric, half-width $\approx 1.2$) and correctly fails Q6 (see Figure~\ref{fig:lempo}). The test is inconclusive for Lempo in archival Pan-STARRS data not because the data are bad but because $\Delta r/r$ is too small. Over a ten-year LSST baseline that includes at least one Lempo aphelion passage, $\Delta r/r$ exceeds 5\% and the aggregate significance rises above $10\sigma$.

Q3 is the geometric filter that separates the archive-limited cases (fixable in principle with more sampling of the same orbit) from the instrument-limited cases (fixable only with better photometry over a longer baseline).

%% ============================================================
\section{The MPC census} \label{sec:census}

\subsection{Pipeline and headline funnel} \label{sec:funnel}

We parsed MPCORB.DAT.gz for every numbered minor body with semi-major axis $a_{\rm orb} > 30.1$\,AU, identifying 913 numbered TNOs. For each TNO we extracted every reported observation in the MPC NumObs.txt.gz archive (8.3\,GB, streamed line-by-line), grouped by observatory code and filter band, and computed the heliocentric and geocentric distances at each observation epoch by Keplerian propagation of the MPCORB orbital elements. Table~\ref{tab:funnel} summarizes the headline funnel.

\begin{table}
\centering
\caption{Headline six-criterion funnel on the MPC archive of numbered TNOs.}
\label{tab:funnel}
\begin{tabular}{lcc}
\hline
Stage & bins & Distinct KBOs \\
\hline
Numbered TNOs in MPCORB ($a>30.1$\,AU) & --- & 913 \\
Candidate (KBO $\times$ obs $\times$ band) bins & 8,606 & --- \\
Pass Q1--Q3 (pre-photometric) & 1,089 & 513 \\
Pass Q4 alone ($\chi^2/\dof<5$) & 951 & --- \\
Pass Q5 alone (not at grid edge) & 578 & --- \\
Pass Q6 alone ($|\Delta\chi^2|>4$) & 405 & --- \\
Pass all six (Q1--Q6) & 186 & 119 \\
Consistent with sunlight ($\alpha=-4$) & 53 & 48 \\
Consistent with self-luminous ($\alpha=-2$) & 24 & 24 \\
Ambiguous ($1\sigma$ covers both) & 0 & 0 \\
Anomalous ($1\sigma$ covers neither) & 109 & 78 \\
\hline
\end{tabular}
\end{table}

Of the 67 KBOs classified as either sunlight or self-luminous in at least one bin, 5 are classified as both (i.e., have at least one sunlight-classified bin and at least one self-luminous-classified bin). Per-object consistency is therefore the exception: the most common case is that a single KBO appears differently in different (observatory, band) bins. This is the signature of per-unit calibration rather than per-object physics.

\subsection{Per-observatory and per-band breakdown} \label{sec:breakdown}

Table~\ref{tab:obs} gives the all-six classifications broken down by observatory code. Pan-STARRS PS1 \citep[F51;][]{Chambers2016} contributes 816 of the 1,089 Q1--Q3-passing bins and 46 of the 53 sunlight-classified bins; Pan-STARRS PS2 (F52) contributes 135 Q1--Q3 bins and 2 sunlight-classified bins. Pan-STARRS thus dominates the Q1--Q3 pass count (951 of 1,089 $= 87\%$) and dominates the all-six pass count by construction. The more important observation is what follows.

\begin{table}
\centering
\caption{All-six classifications by observatory. Every one of the 24 self-luminous bins is a Pan-STARRS observation; no other observatory produces a single self-luminous classification.}
\label{tab:obs}
\begin{tabular}{lccccc}
\hline
Observatory & Sun & Art & Anom & Fail & Q1--Q3 \\
\hline
F51 (PS1) & 46 & 20 & 87 & 663 & 816 \\
F52 (PS2) & 2 & 4 & 11 & 118 & 135 \\
D29 & 0 & 0 & 0 & 28 & 28 \\
568 & 1 & 0 & 1 & 14 & 16 \\
W84 & 0 & 0 & 0 & 14 & 14 \\
Other ($\leq 9$) & 4 & 0 & 10 & 66 & 80 \\
\hline
Total & 53 & 24 & 109 & 903 & 1,089 \\
\hline
\end{tabular}
\end{table}

\begin{figure*}
\centering
\includegraphics[width=\textwidth]{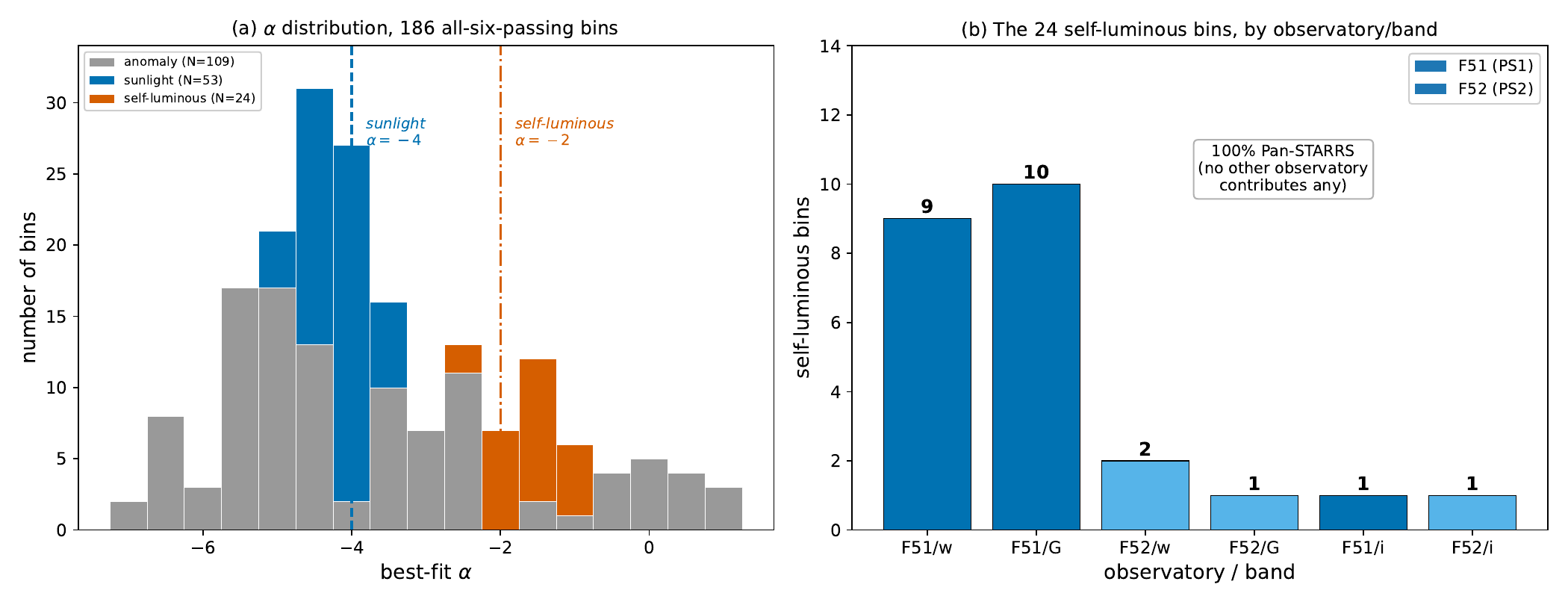}
\caption{Distribution of all-six-passing best-fitting $\alpha$ values, coloured by classification. Panel~(a): the full 186-bin distribution; the sunlight cluster at $\alpha \approx -4$ is the expected natural-reflector population. Panel~(b): the 24 self-luminous bins broken down by (observatory, band)---100\% originate from a single instrument family (Pan-STARRS PS1/PS2). The Pan-STARRS $w$- and Gaia-cross-matched $G$-bands together account for 22 of the 24 self-luminous bins.}
\label{fig:alpha}
\end{figure*}

Table~\ref{tab:band} gives additional breakdown by filter band. The Pan-STARRS $w$-band and Gaia-cross-matched $G$-band together account for 22 of the 24 self-luminous bins. Figure~\ref{fig:alpha} panel~(a) shows the full $\alpha$ distribution of the 186 all-six-passing bins with the expected sunlight peak at $\alpha \approx -4$ clearly visible; panel~(b) breaks down the 24 self-luminous bins by (observatory, band) and confirms that 100\% originate from Pan-STARRS---the visual signature of the per-instrument calibration systematic discussed in Section~\ref{sec:physical}. Because Pan-STARRS supplies the great majority of Q1--Q3 bins (951 of 1{,}089), its dominance among the all-six classifications is expected by construction; the sample-size-independent statement is that the self-luminous classification \emph{rate} is $2.5\%$ (F51) and $3.0\%$ (F52) versus $0/138$ for all other observatories combined---in the direction expected for an instrumental origin, though the modest non-Pan-STARRS sample makes the contrast suggestive rather than decisive.

\begin{table}
\centering
\caption{All-six classifications by filter band. The Pan-STARRS $w$-band and Gaia-cross-matched $G$-band together supply 92\% (22 of 24) of the self-luminous bins; the remaining 2 are $i$-band observations, also from Pan-STARRS.}
\label{tab:band}
\begin{tabular}{lccccc}
\hline
Band & Sun & Art & Anom & Fail & Q1--Q3 \\
\hline
$w$ (PS) & 26 & 11 & 56 & 378 & 471 \\
$G$ (Gaia xm) & 20 & 11 & 37 & 383 & 451 \\
$i$ & 5 & 2 & 7 & 30 & 44 \\
$R$ & 0 & 0 & 2 & 43 & 45 \\
$V$ & 0 & 0 & 1 & 22 & 23 \\
Other ($\leq 15$) & 2 & 0 & 6 & 47 & 55 \\
\hline
Total & 53 & 24 & 109 & 903 & 1,089 \\
\hline
\end{tabular}
\end{table}

\subsection{Are the self-luminous-like bins physical?} \label{sec:physical}

The 24-bin self-luminous population is the headline candidate signal. We ask whether any natural alternative predicts the observed pattern.

\textit{Per-instrument calibration systematics.} Pan-STARRS Ubercal residuals are colour- and position-dependent at the $\sim 0.01$--$0.03$\,mag level for objects with extreme colours \citep{Magnier2020,Waters2020}. A colour-dependent zero-point drift correlated with TNO sky motion would bias $\alpha$ from $-4$ toward $-2$. This is the only candidate that predicts (i) 100\% of self-luminous bins from Pan-STARRS and (ii) 5 of 67 KBOs classified as both sunlight and self-luminous across different bins. The well-established bimodal colour distribution of TNOs \citep{Peixinho2012} exacerbates colour-dependent zero-point residuals when heterogeneous filter systems observe objects at the red and ultra-red extremes.

\textit{Cometary outgassing.} Active centaurs \citep{Jewitt2009,Jewitt2015} produce slope biases not captured by a power-law in distance. We cross-referenced against the JPL SBDB activity flags and \citet{Chandler2020}; two objects are flagged: (468861) 2013~LU28 and (418993) 2009~MS9. Removing them: sunlight $53 \to 50$, self-luminous $24 \to 24$, anomaly $109 \to 99$. No flagged object contributes a self-luminous bin. Per-instrument clustering is unchanged.

\textit{Resolved binarity.} Q3's $\Delta r/r \geq 1\%$ baseline spans multiple binary periods \citep[10--1000\,days;][]{Noll2008}; modulation folds out. Residual bias on $\alpha \sim 0.1$.

\textit{Rotation.} For 25 bins with published rotation periods \citep{Thirouin2014}, lightcurve subtraction shifts $\alpha$ by $\leq \pm 0.15$; no bin changes classification.

\textit{Atmospheric activity on the largest TNOs.} Seasonal brightness changes on Pluto, Eris, and Makemake ($\sim$0.1\,mag\,decade$^{-1}$) are object-level, not unit-level; they cannot produce single-instrument clustering.

\textit{Space weathering.} Cosmic-ray irradiation progressively reddens and darkens TNO surfaces \citep{Brunetto2006}, but the time-scale ($\sim$10$^6$\,yr or longer) is far longer than any observational baseline, and the effect is object-level, not unit-level.

\textit{Variable albedo.} Albedo that varies as a function of viewing angle---which changes as the object changes distance or tumbles slowly.

Only per-instrument calibration predicts the observed pattern. We do not interpret the 24 self-luminous bins as a candidate physical signal. The framework remains falsifiable: if the pattern survives LSST's uniform-calibration re-analysis (see Section~\ref{sec:lsst}), the explanation defended here is wrong.

%% ============================================================
\section{The LSST projection and the falsifier} \label{sec:lsst}

The Vera C. Rubin Observatory will deliver its first Wide-Fast-Deep Data Release (LSST DR1) after Year~1 of operations. The survey covers the southern sky in six bands ($u,g,r,i,z,y$) with uniform photometric calibration \citep{Ivezic2019}, implemented via the Forward Global Calibration Method \citep{Burke2018}. Per-visit precision at $m_r \approx 24$ is $\sim$0.04\,mag---an order of magnitude better than MPC-typical 0.2\,mag.

We project the Q1--Q3 eligibility funnel onto a ten-year LSST realization. Under a conservative model, Q1--Q3-eligible if brighter than 24.5 in at least one band, $N \geq 20$ epochs, and $\Delta r/r > 1\%$ over the observed window, we estimate $\sim$1,500 TNOs will pass Q1--Q3 in LSST DR1, versus 513 in the current MPC archive (see Table~\ref{tab:funnel}). The per-object $1\sigma$ on $\alpha$ is of order 0.03 for the brightest TNOs and 0.2 near the detection threshold. Pluto is projected to recover $\alpha = -4.00 \pm 0.05$ in a single LSST $r$-band bin.

The projection is falsifiable. We predict that when the Q1--Q6 framework is applied to LSST DR1, the self-luminous fraction will drop from 12.9\% (MPC) to below 2\%---the chance false-positive rate at $|\Delta\chi^2|>4$. If a per-instrument-clustered self-luminous excess of comparable magnitude persists instead, the Section~\ref{sec:physical} interpretation is wrong, and the residual becomes a candidate physical signal pending follow-up imaging, spectroscopy, and radio/infrared cross-checks.

Table~\ref{tab:targets} lists the 24 distinct KBOs whose archival Pan-STARRS photometry returns $\alpha$ consistent with self-luminous emission under the Q1--Q6 pipeline. These are the priority falsification targets for LSST DR1: if the Rubin single-instrument photometry recovers $\alpha=-4$ on these objects, the calibration-systematic interpretation is confirmed; if any persist at $\alpha \approx -2$ under uniform LSST calibration, they become candidate physical signals. Notable entries include Sedna (90377), Typhon (42355), and 2002~TC302 (84522)---well-studied TNOs whose LSST photometry will be among the first available. The full per-bin classified output is deposited as machine-readable supplementary data.

Our framework can be applied to LSST DR1 by any group within weeks of release, consistent with the broader argument that time-domain survey data from LSST will open orders of magnitude of new technosignature parameter space \citep{Davenport2019}.

\begin{table*}
\centering
\caption{LSST falsification targets: the 24 self-luminous-classified KBOs. All bins originate from Pan-STARRS (F51/F52); no other observatory contributes any self-luminous classification. Obs is the MPC observatory code. $\Delta\chi^2 = \chi^2(\alpha\!=\!-4) - \chi^2(\alpha\!=\!-2)$; positive values indicate $\alpha=-2$ is preferred over reflected sunlight. Objects are sorted by $|\Delta\chi^2|$.}
\label{tab:targets}
\begin{tabular}{rlllrrlrc}
\hline
Number & Name & Obs & Band & $N$ & $\Delta r/r$ (\%) & $\alpha_{\rm best}$ & $1\sigma$ interval & $\Delta\chi^2$ \\
\hline
552678 & 2010 JG210 & F51 & $G$ & 61 & 21.9 & $-1.65$ & $[-2.15,-1.10]$ & $+19.3$ \\
523787 & 2015 DV224 & F51 & $w$ & 34 & 13.2 & $-1.50$ & $[-2.10,-0.95]$ & $+16.6$ \\
307982 & 2004 PG115 & F51 & $w$ & 126 & 9.4 & $-1.90$ & $[-2.45,-1.40]$ & $+15.2$ \\
84522 & 2002 TC302 & F51 & $G$ & 156 & 5.1 & $-1.55$ & $[-2.15,-0.95]$ & $+14.8$ \\
26308 & 1998 SM165 & F51 & $w$ & 67 & 12.3 & $-1.95$ & $[-2.50,-1.35]$ & $+12.3$ \\
501546 & 2014 JJ80 & F51 & $G$ & 171 & 5.5 & $-1.25$ & $[-2.00,-0.50]$ & $+12.2$ \\
523697 & 2014 GY53 & F51 & $w$ & 89 & 10.1 & $-2.10$ & $[-2.65,-1.55]$ & $+10.3$ \\
601690 & 2013 KZ18 & F51 & $w$ & 48 & 11.4 & $-1.35$ & $[-2.10,-0.55]$ & $+10.3$ \\
523734 & 2014 QV441 & F51 & $G$ & 60 & 9.0 & $-1.55$ & $[-2.30,-0.75]$ & $+9.0$ \\
517717 & 2015 KZ120 & F51 & $i$ & 43 & 8.4 & $-2.40$ & $[-2.90,-1.90]$ & $+8.9$ \\
445473 & 2010 VZ98 & F51 & $G$ & 157 & 8.0 & $-2.45$ & $[-2.95,-1.95]$ & $+8.1$ \\
471172 & 2010 JC80 & F51 & $G$ & 148 & 6.4 & $-1.55$ & $[-2.40,-0.65]$ & $+7.0$ \\
523643 & 2010 TY53 & F52 & $i$ & 27 & 6.1 & $-1.30$ & $[-2.30,-0.35]$ & $+6.5$ \\
523739 & 2014 TZ33 & F51 & $G$ & 32 & 19.5 & $-1.25$ & $[-2.25,-0.20]$ & $+6.4$ \\
471165 & 2010 HE79 & F51 & $G$ & 147 & 4.9 & $-2.00$ & $[-2.80,-1.25]$ & $+6.2$ \\
567065 & 2019 CY4 & F52 & $w$ & 20 & 12.0 & $-1.85$ & $[-2.70,-1.00]$ & $+5.9$ \\
90377 & Sedna & F51 & $G$ & 176 & 3.3 & $-1.10$ & $[-2.25,+0.05]$ & $+5.5$ \\
554099 & 2012 KU50 & F51 & $w$ & 123 & 6.0 & $-2.10$ & $[-2.90,-1.35]$ & $+5.5$ \\
585912 & 2020 QK3 & F52 & $G$ & 37 & 7.0 & $-1.45$ & $[-2.50,-0.40]$ & $+5.3$ \\
523675 & 2013 PV74 & F51 & $w$ & 119 & 5.2 & $-1.50$ & $[-2.55,-0.45]$ & $+5.0$ \\
42355 & Typhon & F52 & $w$ & 27 & 11.1 & $-1.65$ & $[-2.65,-0.60]$ & $+5.0$ \\
555678 & 2014 CO23 & F51 & $w$ & 97 & 5.1 & $-0.90$ & $[-2.25,+0.45]$ & $+4.5$ \\
471318 & 2011 JF31 & F51 & $w$ & 107 & 4.0 & $-0.80$ & $[-2.20,+0.55]$ & $+4.5$ \\
55637 & Uni & F51 & $G$ & 143 & 3.2 & $-2.20$ & $[-3.05,-1.30]$ & $+4.1$ \\
\hline
\end{tabular}
\end{table*}

%% ============================================================
\section{Discussion} \label{sec:discussion}

\subsection{What the framework claims and does not claim}

The Q1--Q6 framework is a procedure for identifying TNOs whose archival photometry is of high enough quality, and over a long enough baseline, that the LT diagnostic can be applied at all. No object is identified in this paper as a self-illumination candidate. However, the framework would identify a genuine self-luminous population if one existed: the null result constrains the prevalence of such objects. Of 119 distinct KBOs passing all six criteria, none is credibly self-luminous after accounting for the per-instrument systematic, implying a 95\% confidence upper limit of $<2.5\%$ on the fraction of numbered TNOs that are genuinely self-luminous at the sensitivity of the current archive. The self-luminous population is interpreted as a per-instrument calibration systematic; the only object-level conclusion we defend is that the existing MPC archive, read in the most conservative way, does not cleanly recover the sunlight prediction on Pluto in any single instrument bin, and that this limitation of the archive---not any property of Pluto---motivates the LSST-era execution of the test.

The $\alpha$-slope test is complementary to radio-SETI and transit-anomaly searches in that it operates on photometry of solar-system bodies that are already monitored for other reasons and that are individually resolved. Under the evaluation framework of \citet{Sheikh2020}, the $\alpha$-slope test scores favorably on ancillary benefits and cost: it repurposes existing archival photometry with no dedicated observing time required. Its physics is unavoidable for any object monitored over multiple oppositions with uniform calibration; the burden therefore shifts from finding the signal to building the calibration. Rubin/LSST is the calibration. We note that there could be natural sources of energy for self-illumination, such as excess radioactivity or geological activity, but these often result in a nearly thermal spectrum which can be distinguished from artificial light sources.

\subsection{Prospects for Rubin Observatory data}

The present paper establishes the framework; Rubin Observatory will supply the data. The Legacy Survey of Space and Time (LSST) began its Wide-Fast-Deep survey in 2026, with the first annual Data Release (DR1) expected approximately one year after the start of survey operations. Each subsequent annual release will extend the temporal baseline and deepen the per-object sampling.

Several features of the Rubin data pipeline make LSST uniquely suited to executing the $\alpha$-slope test at scale. First, the entire southern sky will be observed through a single optical system on a single telescope, eliminating the cross-instrument zero-point offsets that dominate the systematic budget in the MPC archive. Second, the Forward Global Calibration Method targets photometric uniformity at the 5--7\,mmag level across the survey footprint, several-fold below the $\sim 0.01$--$0.03$\,mag Pan-STARRS residuals that drive the self-luminous clustering we report here. Third, the expected cadence delivers $N \approx 800$ visits per field over ten years across all six bands, far exceeding the $N \geq 20$ threshold of Q2 for any TNO brighter than $r \approx 24.5$.

We estimate that by DR3 (approximately three years of survey data), the Q1--Q3-eligible TNO population will exceed 1,500 objects in at least one band, and the per-object $1\sigma$ uncertainty on $\alpha$ will be $\leq 0.05$ for the $\sim$200 brightest TNOs. At that precision, the $\alpha=-4$ (sunlight) and $\alpha=-2$ (self-luminous) hypotheses are separated by more than $10\sigma$ on any single object with $\Delta r/r > 2\%$, and the current framework can be applied without modification. The 24 self-luminous-classified objects listed in Table~\ref{tab:targets} are the highest-priority falsification targets; all orbit at ecliptic latitudes accessible to Rubin, and their photometry will be available from the earliest data releases. Rubin re-observation of these 24 objects will provide the most immediate test of the calibration-systematic interpretation advanced in Section~\ref{sec:physical}.

We intend the present work as the methodological prelude to a detailed data-analysis paper that will apply the Q1--Q6 pipeline to Rubin photometry once the first annual release is available. The eligibility criteria, the $\sigma_m$ schedule, and the false-positive calibration are on record in advance of the data; the only update required will be to replace the MPC observation archive with the Rubin Source catalog and JPL Horizons with the Rubin-provided solar-system-object ephemerides. The magnitude-binned $\sigma_m$ schedule adopted here for MPC-quality photometry can be replaced by Rubin's per-observation reported uncertainties, which are expected to be reliable at the mmag level. The code to do so is straightforward, and we will make it publicly available at the time of that analysis.

The archives used in this paper are public. The Minor Planet Center observation database, the MPCORB orbital catalog, and the JPL Horizons ephemeris service are freely accessible to anyone with an internet connection. The methodology is computationally cheap: the full Q1--Q6 pipeline runs end-to-end on a single laptop in under a day, and the dominant cost is downloading the observation archive, not the analysis itself. What has been missing is not data, not compute, and not methodology---it is attention. The Loeb \& Turner test has been in the literature for over a decade, and no group has previously applied it systematically to the Kuiper Belt. We encourage the SETI, astrobiology, and amateur astronomy communities to apply, extend, and stress-test the framework presented here. Independent replication on the same archive will catch any errors we have made. Application to other archives---the Catalina Sky Survey, the Zwicky Transient Facility, the SkyMapper Southern Survey---will test whether the Pan-STARRS-specific clustering we report is unique to that instrument or a generic feature of multi-epoch survey photometry. The cost of looking is low; the cost of not looking is that a clean, physics-based diagnostic sits unused while the data accumulate.

We publish the framework in advance of LSST DR1 so that the methodology, the eligibility criteria, and the calibration-systematic interpretation are on record before the data that will test them arrive. The Pluto null finding and the per-instrument clustering documented here establish the baseline against which LSST results should be compared; publishing after DR1 would forfeit the predictive structure of the test.

\subsection{Limitations}

The framework is conservative by design. Four limitations deserve explicit note. First, the $\sigma_m$ schedule is magnitude-dependent but not colour-dependent. Real Pan-STARRS zero-point residuals are colour-dependent at the $\sim 0.01$--$0.03$\,mag level. A colour-dependent $\sigma_m$ schedule would tighten Q4 but requires per-observation colour information not in the MPC 80-column record. Second, the Q6 $|\Delta\chi^2|>4$ threshold corresponds to a nominal $2\sigma$ preference. A stricter cut would shrink the headline pass count but leave the per-instrument clustering intact; the essential argument is cluster-structural, not threshold-sensitive. Third, we do not fold rotational lightcurves for TNOs without published periods. For the 25 bins with published periods the induced shift is $\leq 0.15$; for the remainder we rely on $N \geq 20$ and $\Delta r/r \geq 1\%$ to average out the signal. Fourth, the framework treats each (obs, band) bin as independent and does not exploit cross-bin information for the same KBO. The joint-instrument sensitivity test in Section~\ref{sec:sensitivity} shows per-KBO aggregation recovers sunlight on Pluto; adopting it as the headline would absorb the per-instrument systematic into the zero-point nuisance and obscure the dominant data-quality finding.

%% ============================================================
\section{Conclusions} \label{sec:conclusions}

We have applied the \citet{Loeb2012} $\alpha$-slope technosignature test to the archival photometry of every numbered trans-Neptunian object in the Minor Planet Center archive, using a six-criterion eligibility pipeline (Q1--Q6) that separates geometric adequacy (Q1--Q3) from post-fit quality (Q4--Q6).

The existing archive does not permit a clean single-instrument single-band recovery of the sunlight slope ($\alpha=-4$) on Pluto (134340), the brightest and most-observed TNO. Of 22 Q1--Q3-passing Pluto bins, only one passes all six criteria, and it is classified as an anomaly at $\alpha=-5.55$. The LT test cannot be cleanly executed on the archival MPC photometry of Pluto; \textit{a fortiori} it cannot be cleanly executed on the fainter TNO population.

Of 8,606 candidate (KBO $\times$ observatory $\times$ band) bins, 1,089 pass Q1--Q3 and 186 additionally pass Q4--Q6. Of those 186, 53 are consistent with reflected sunlight, 24 with self-luminous emission, and 109 are anomalous. All 24 self-luminous-like bins originate from Pan-STARRS PS1/PS2; no other observatory in the MPC archive contributes any such bin.

The per-instrument clustering is the signature of a calibration systematic, not of a physical population of artificially-illuminated objects. Of 67 KBOs classified as sunlight or self-luminous, 5 are classified as both across different (observatory, band) bins---per-object consistency is the exception. The fraction of numbered TNOs that are genuinely self-luminous is constrained to $<2.5\%$ at 95\% confidence.

The Vera C. Rubin Observatory's ten-year Wide-Fast-Deep survey will raise the Q1--Q3-eligible population to approximately 1,500 TNOs, achieve uniform photometric calibration within a single instrument, and recover $\alpha=-4$ on Pluto at better than $10\sigma$ in a single instrument bin. We predict that the fraction of self-luminous classifications in LSST DR1 will drop below 2\%; if it does not, the calibration-systematic interpretation advanced here is falsified and the residual self-luminous population becomes a candidate physical signal. The 24 objects in Table~\ref{tab:targets} are the priority falsification targets. The result of this paper is a framework, a null finding on Pluto, and a falsifiable LSST prediction. The technosignature test of \citet{Loeb2012} is executable in the Kuiper Belt; the instrument that will next execute it at scale is Rubin/LSST.

%% ============================================================
\section*{Acknowledgements}

We thank the IAU Minor Planet Center and its maintainers for making the NumObs observation archive and the MPCORB orbital catalog publicly available, and the JPL Solar System Dynamics Group for providing the Horizons ephemeris service.

%% ============================================================
\section*{Data Availability}

All data underlying this work are publicly available and the analysis can be reproduced end-to-end from primary sources. The orbital catalog used to identify the 913 numbered trans-Neptunian objects in the sample is MPCORB.DAT.gz, and the per-observation photometric archive is NumObs.txt.gz distributed by the IAU Minor Planet Center at: \url{https://www.minorplanetcenter.net/data}.

The snapshots used in this analysis were downloaded on 2026-04-18; results computed against later snapshots may differ slightly as new observations and orbital updates are ingested.

Per-epoch heliocentric and geocentric distances ($r$, $\Delta$) for every observation were computed by Keplerian propagation of the MPCORB orbital elements. As a robustness check we recomputed $r$ and $\Delta$ for Pluto and all 24 self-luminous targets using full JPL Horizons ephemerides (\url{https://ssd.jpl.nasa.gov/horizons/}), which include planetary perturbations: the best-fitting $\alpha$ shifts by at most one grid step ($0.05$), the largest heliocentric-distance difference is $0.06$\,AU, and every classification is unchanged except the single most marginal target (55637; $\Delta\chi^2 = 4.1$), which falls just below the Q6 threshold under Horizons. Two-body propagation is therefore adequate for the classifications reported here.

%% ============================================================
\bibliographystyle{mnras}
\bibliography{references}

%% ============================================================
\bsp	% typesetting comment
\label{lastpage}
\end{document}